# Low temperature conductivity of $BaFe_{0.5}Nb_{0.5}O_3$ double perovskite structure ceramics

Vijay Khopkar and Balaram Sahoo†

*Materials Research Centre, Indian Institute of Science, Bangalore 560012 India*

†Corresponding author: bsahoo@iisc.ac.in (B. Sahoo)

Ph: +91 8022932943

# Abstract

In this work, we explore the origin and the type of charge carries, and their transport mechanisms in polycrystalline barium-iron-niobate (BFN, $BaFe_{0.5}Nb_{0.5}O_3$) ceramics, at lower temperatures between 20 K and 300 K. The observed point defects at grain and surface defects at grain boundary region are responsible for the electronic conductivity, whereas dipoles of the grain region are responsible for the ionic conductivity, these independent electronic and ionic conductivity were responsible for the total conductivity of our BFN sample. The required activation energy for conduction of electrons in grain boundary and ions in grain region were calculated to be 317 meV and 17 meV respectively. The electronic conductivity of grain region obeys Jonscher's universal power law. Analysis of the temperature dependent frequency exponent suggests that the electronic conductivity of grain follows the overlapping large polaron tunneling (OLP) model which further validated. The temperature dependent conductivity follows the Mott variable range hopping (VRH) model, showing hopping range increase with decreasing temperature. The defect density of the grain obtained to be $3.17 \times 10^{17}$ $eV^{-1}cm^{-3}$. The contribution of phonons to the conductivity understood by considering Schnakenberg model, Activation energy of 110 meV corresponds to the multiphoton (both optical and acoustic) and 1.6 meV corresponds to the acoustic phono only. Our systematic study provides in-depth understanding of the low temperature conductivity mechanism of polycrystalline BFN ceramics.

**Keywords:** Barium-iron-niobate; conductivity; Charge carriers; Low temperature conduction mechanism; Overlapping-large-polaron-tunneling; Mott variable range hopping.

## 1 : Introduction

Materials with high dielectric constant find their applications in diverse areas, such as in mechanical sensors, actuators, energy storage devices, memory devices and gas sensors, etc.[1,2]. For this purpose, lead based $ABO_3$ type perovskite structured materials were traditionally looked at[3–5]. But, due to the toxic nature of lead (Pb), attention is shifted towards finding, alternate, lead-free perovskite materials[6]. In this regard, many perovskites, such as $BaTiO_3$, $BiFeO_3$ and $KNbO_3$ etc. and their compositional modifications are studies in detail[7,8,9]. Generally, the simple materials ($BaTiO_3$, $BiFeO_3$ and $KNbO_3$ etc.) are ferroelectric in nature and they show high dielectric constant only at the ferroelectric-paraelectric phase transition temperature (Curie temperature). However, for application purposes it is desirable to select materials having high dielectric constant over a wide range of working temperatures. Therefore, much attention is given to modify the perovskite materials with the expectation to induce a diffused phase transition (DPT)[10]. The idea here is to create a distribution of local regions having different phase transition temperatures. The exact strategy that can be used to induce a DPT in pervoskite materials, is by suitably choosing two or more cations for the A and B – site cations and by introducing a spatial and compositional disorder in their occupancies [11,12]. With increase in this disorder, most of the materials begin to exhibit diffuse phase transition, i.e., the temperature of transition from the ferroelectric to paraelectric state can vary over a large temperature range, supporting the dielectric constant to remain high over a wide range of temperatures. For this reason, a partial substitution of a different cation at the A or B site of the pervoskite oxides, with the general formula $A'A''B_2O_6$ (such as $K_{0.5}Na_{0.5}NbO_3$) or $A_2B'B''O_6$ (such as $BaFe_{0.5}Nb_{0.5}O_3$) gained interest. Such perovskites are often called double perovskites. In addition to the DPT, these double perovskite materials also show various other interesting properties and are suitable for many different application[9].

Barium iron niobate (BFN), $BaFe_{0.5}Nb_{0.5}O_3$, is an interesting lead-free double pervoskite material having a two different cations ($Fe^{3+}$ and $Nb^{5+}$) occupying at the B site[13]. Saha and Sinha, reported that BFN is a lead free relaxor ferroelectric material [14,15]. The high value of dielectric constant observed in this material (BFN) originates due to the Maxwell – Wagner polarization[16,17]. The dc conductivity measurements performed between room temperature (RT) to 550 ºC show that electrical transport mechanism in BFN follows a jump relaxation model[18]. Bhagat and Prasad also studied the high temperature dielectric and conductivity behavior of BFN. They observed that, at this high temperature, the charge transport mechanism can be successfully explained by the correlated barrier hopping (CBH) model[19]. All these studies support the Maxwell-Wagner type polarization mechanism in BFN sample at high temperature, which leads to the high polarization values at lower frequencies.

To understand the polarization mechanism and its origin accurately, it is necessary to identify the type of charge carriers responsible for the conductivity in BFN and also to discern the conductivity emanating from the grains and the grain boundaries of the polycrystalline ceramic samples. This requires a temperature dependent conductivity study down to a low temperatures of ~ 10-20 K, because at this low temperature, the ac and/or dc conductivity of grains and grain boundary regions is expected to be different. Along with the temperature dependent conductivity behavior, this low temperature study can reveal the quality of the materials. Because, the contribution of small amount of grain boundaries would be less if the material is dense and having bigger grains. However, in the literature, only one report is available on the lower temperature conductivity study of this material, but that was still not down to very low temperature. In that report, Saha and Sinha studied the ac conductivity at medium-low temperatures between 93 and 213 K [14]. As they measured down only to 93 K, this study does not provide any systematic understanding of the polarization and conductivity mechanism in BFN. Hence, the obtained information is limited and it does not providing any clarity on the mechanism of low-temperature ac and dc conductivity mechanism in BFN[14]. As per our knowledge, apart from this medium-low temperature study, no detailed systematic study down to a very low temperature (of ~ 10-20 K) on the electrical conductivity mechanism in BFN is available. Hence, in order to explore the usefulness of BFN and related materials for various optical and electrical applications, a microscopic understanding of the charge transport process is necessary. In this work, we systematically studied the conductivity mechanism of BFN ceramic sample at low temperatures between 20 and 300 K, and explored the charge transport mechanism.

## 2 : Experimental details

Polycrystalline $BaFe_{0.5}Nb_{0.5}O_3$ sample is synthesized by a conventional solid state technique, using $Nb_2O_5$ (99.95%, Spectrochem), $Fe_2O_3$ (95%, SDFCL) and $BaCO_3$ (99%, Merck) as precursors. For the synthesis, stoichiometric amounts of the precursors were initially mixed and ground using agate mortar and pestle, to get a homogeneous mixture. The mixture was then calcined at optimized conditions, 1200 °C for 4 h, to obtain phase pure BFN sample. By mixing 3 wt % polyvinyl alcohol (as a binder) with the phase pure powder, disc-shaped pellets (with 10.34 mm diameter and 1.02 mm thickness) were prepared using a uniaxial press (load applied: 25 kN). The pellets are then sintered at 1200 °C for 2 hours in air. A dense pellet was then polished and attached with copper electrodes using low temperature conductive silver paste (Siltech inc.). This (pellet) sample was used for the low temperature dielectric measurement. The dielectric measurement was performed using precision impedance analyzer (4294A, Agilent) in a frequency range of 40 Hz to 5 MHz with an applied voltage of 500 mV. Measurements were carried out between 20 K and 300 K in steps of 5 K using close cycle refrigerator (CH-204SFF, JANIS) with Helium compressor (HC-4E, Sumitomo). For the measurements the temperature was ramped at a rate of 5 K/min. with a dwelling time of 5 min. The obtained dielectric data were used to obtain the conductivity at different frequencies and at different temperatures.

The powder X-ray diffraction (XRD) pattern of a crushed pellet sample was obtained by "PANalytical X'PertPro" diffractometer (Cu K$_\alpha$ radiation ($\lambda$ = 1.54 Å) and Ni filter). The XRD pattern was recorded in the 2θ range of 10 - 90° with a step size of 0.0263°. Scanning electron microscopy (SEM) images were obtained through ′ESEM Quanta′ instrument at an operating voltage of 10 kV. Transmission electron microscopy (TEM) images were obtained by JEOL JEM - 2100F microscope. The high resolution TEM (HRTEM) images were analyzed with ′digital micrograph′ software. The density of the sample was measured by Archimedes' principle using xylene (density = 0.87 g/cm$^{-1}$) as the liquid medium, in a homemade apparatus [13]. The UV-Vis spectra were measured using the LAMBA 750 UV/Vis/NIR spectrometer from PerkinElmer with deuterium and tungsten halogen lamp as the light source, and a PbS detector. The XPS data collected by Kratos Axis Ultra with Al K$\alpha$ source (1845.6 eV) and analyzed by Casa XPS software. All spectra calibrated with adventitious C1s peak having BE value of 284.8 eV along with U 2 Tougaard type background. Survey scan analyzed to determine the stoichiometry of the compound and presence of adventitious carbon and contamination.

## 3 : Result and discussion

### 3.1 : Structure and microstructure of the BFN sample

Fig. 1(a) shows the powder XRD pattern of the synthesized BFN sample obtained after calcination of the ground precursor powder at 1200 °C for 4 h. From XRD powder pattern it is confrment that phase pure BFN (with space group: *Pm-3m*) is formed. The SEM image (Fig. 1(b)) of the pellet shows distinct grains and grain boundaries. The mass density of the pellet used for conductivty measurement was measured to be high (i.e., 96 % of the theoretical density of BFN).

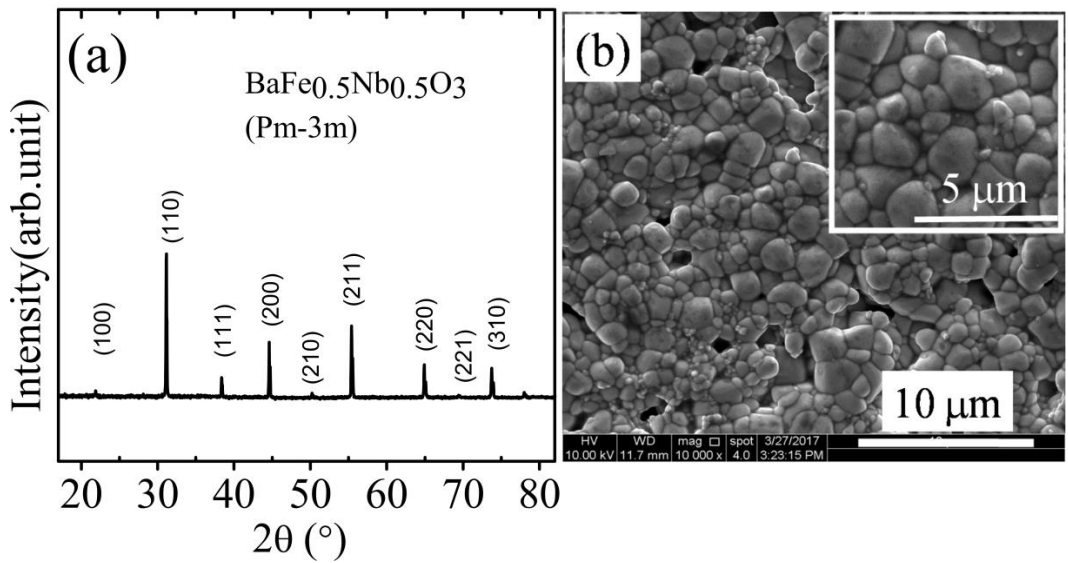

Fig. 1(a) The powder XRD pattern, and (b) the SEM image of BFN ceramics.

Fig.2 (a) and (b) shows TEM and selected area electron diffraction (SAED) pattern of BFN particles. Observed reflections are (110), (111), (200) and (211) plains with corresponding d spacing values shown in Fig.2 (b). Intensity of all these reflections are matches with intensity observed in XRD [Fig. 1(a)], which further confirm the formation of our phase pure BFN sample. Fig.2 (c) and (d) shows respectively HRTEM image and SAED pattern of highlighted region of Fig.2 (a). The calculated value of d spacing 2.95 Å showing presence (110) lattice plain. Further the higher order reflections of 110 lattice plane were indexed as shown in (d). This shows the highly crystalline nature of our BFN powder sample. Fig.2 (e) shows the grain (G) and grain boundary (GB) region of BFN particle. Further Fig.2 (f) and (g) shows the surface and point defects (interstitial defects) observed in grain boundary and grain region respectively. From the TEM study it is observed that the BFN particles made up of many grains separated by distinct grain boundaries. Our results clearly shows that the ions are well ordered within the grain regions with point defects (interstitial) and grain boundary regions show disordered atomic arrangement (surface defect). These disorder (defects) produces free and/or trap charge carriers which are responsible for conductivity. The origin of these charge carries due to the defects are represented by Kroger –Vink notation in the section 3.8. The polycrystalline nature of our BFN ceramic sample and high temperature sintering mainly responsible for observed various types of observed defects[20].

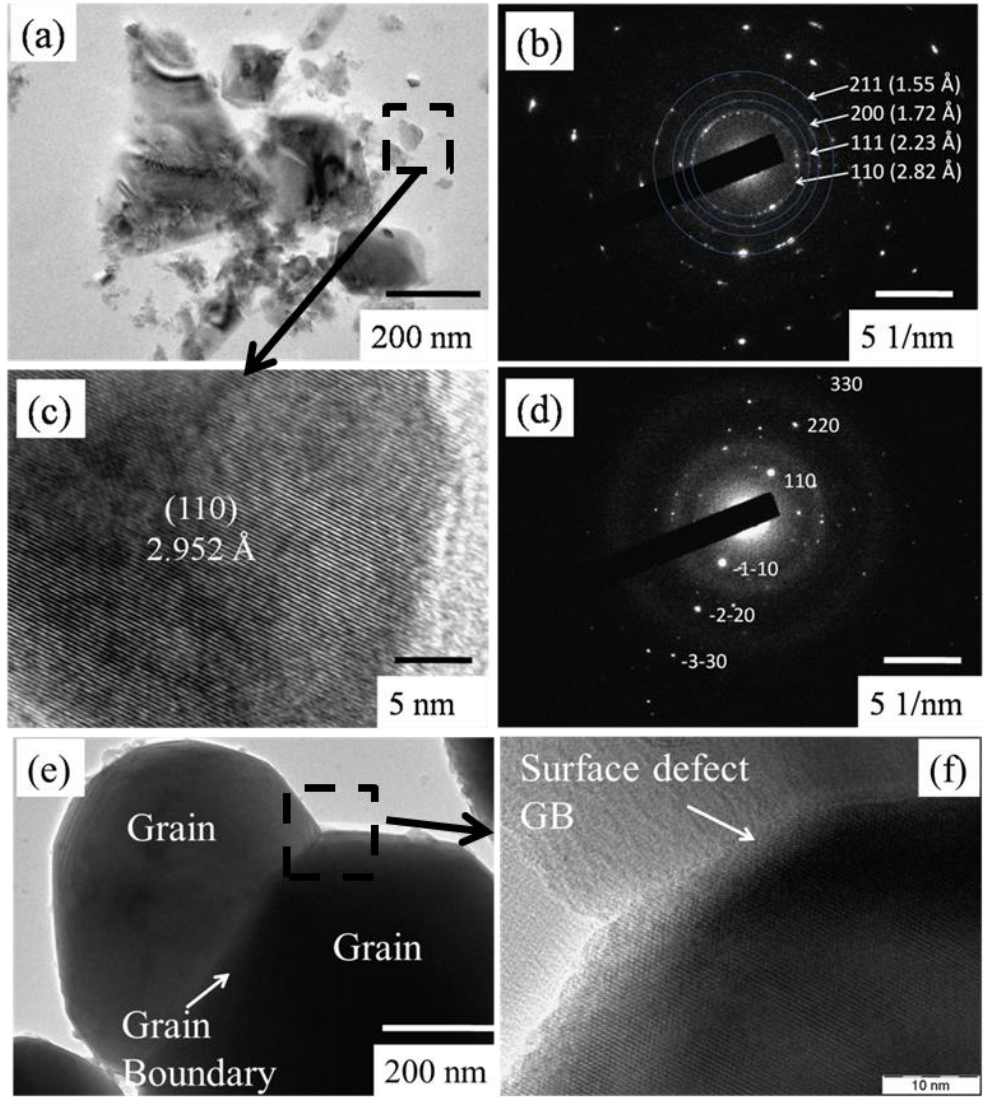

Fig. 2 (a, b) TEM image and SAED pattern of polycrystalline particles, (c, d) HRTEM and SAED pattern of crystalline grain (e) grain and grain boundary (GB) region (f) surface and defects of BFN ceramic sample.

## 3.2 : Optical property of BFN ceramic

To calculate the energy band gap and presence of defect energy levels of our BFN sample, UV-Vis diffuse reflectance spectroscopy (DRS) measurement used. According to the theoretical model of optics of coating given by the Kubelka and Munk[21,22], the Kubelka – Munk (K-M) function give as,

$$F(R_\infty) \equiv \frac{K}{S} = \frac{(1-R_\infty)^2}{2R_\infty} \quad (1)$$

Here $R_\infty$ is the measured diffusion reflectance, $K$ and $S$ is the absorption and scattering constant respectively. K-M function (Eq.1) gives the relation between diffusion reflectance ($R_\infty$) and optical loss ($L = K+S$). Assume that, for all incident radiations scattering is constant ($S$ constant) and incident radiation perfectly scattered in the diffusive manner then, $K$ becomes equal to $2\alpha$. Hence Eq. (1) becomes, $F(R_\infty) = C_1 \alpha$ giving relation between experimentally measured diffusion reflectance (or K-M function) with the absorption coefficient. Here, $C_1$ and $\alpha$ is the constant and absorption coefficient respectively. For 3D solid, this absorption coefficient for incident photon of energy $h\nu$ related to the optical band gap ($E_g$) and density of states at conduction and valence band. The relation is given by the Tauc's equation as[23–25],

$$\alpha h\nu = C_2 (h\nu - E_g)^n \quad (2)$$

Since Eq. (1) becomes, $(\alpha h\nu)^{1/n} = [F(R_\infty)h\nu]^{1/n} = C_3 (h\nu - E_g) \quad (3)$

Here, $n$ can take values ½ and 2 for direct and indirect transition respectively. The Urbach energy ($Eu$) associate with defects was calculated by using Urbach equation as[25,26],

$$\alpha = \alpha_0 \exp(h\nu / E_u) \quad (4)$$

α₀ is constant. Fig.3 (a) and (b) shows Tauc's plot and Urbach plot of the BFN sample calculated by using Eq. (3) and Eq. (4) respectively. The band gap energy of our BFN sample obtained to be 1.88 eV energy and Urbach energy of 534 meV. The Fig.3 (c) shows the schematic of the band structure of our BFN sample. This shows that energy of 1.88 eV is required for electrons in the valence band to go in the conduction band for electronic conduction. More importantly in our material there observed to be Urbach energy 534 meV, associated with intrinsic defects (disorder) [Fig.2 (e) and (f)].

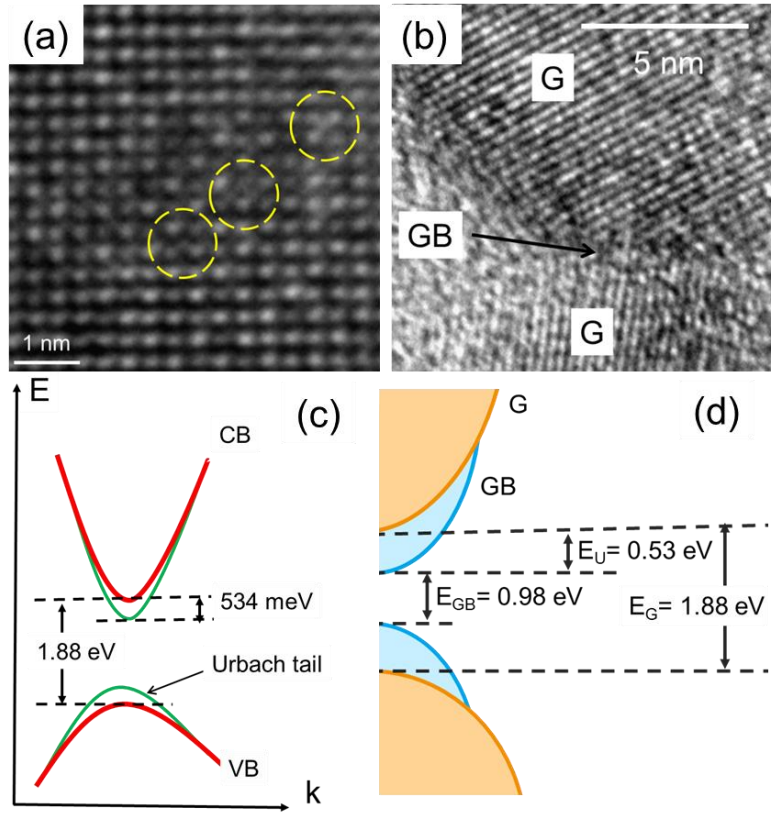

Fig.3 (a) point defect (b) surface defect and (c,d) schematics of band structure of BFN (not in scale) with presence of defects level (dash).

From the analysis of HRTEM and UV-Vis spectroscopy measurement is confirmed that our BFN polycrystalline ceramic sample has various kinds of defects. The large disorder at grain boundary region with continues path for conduction [Fig. 2(f)]. Also in the grain, points defects like interstitial [Fig. 2(g)] and vacancies for the localized conduction of charge carrier. In external applied electric field with variable frequency these charge carriers from different regions response differently as discussed in following sections.

**3.3 : Dielectric constant (ε*) and electric conductivity (σ*)**

Recently, the systematic study of the temperature and the frequency dependent dielectric constant, electric modulus and impedance behavior of BFN is reported [13]. For convenience, the real ($\varepsilon'$) and imaginary ($\varepsilon''$) parts of the dielectric constants versus frequency measured at four different temperatures, namely 20, 100, 200 and 300 K, is given in Fig. 4 (a, b). From the analysis of the results, it was clear that there exists contribution from grain (G) and grain boundary (GB) regions to the impedance of the sample. Furthermore to study the conductivity of our BFN sample, these measured dielectric data converted into the conductivity with formulation given as[27–32],

$$\sigma_T^*(\omega) = \sigma_R(\omega) + i\sigma_I(\omega)$$
$$\sigma_I(\omega) = A_I \omega^S = \varepsilon_0 \omega \varepsilon'(\omega)$$
$$\sigma_R(\omega) = \sigma_{dc} + A_R \omega^S = \sigma_{dc} + \varepsilon_0 \omega \varepsilon''(\omega) \quad (5)$$

Where $\sigma_R$ real part $\sigma_I$ imaginary part of conductivity, $\varepsilon_0$ Permittivity of free space, $\omega$ frequency, $\sigma_{dc}$ dc conductivity, $A$ complex constant and $S$ frequency exponent $[S = (d\ln\sigma)/(d\ln\omega)]$. The Eq. (5) is the 'Universal dielectric relaxation' (UDR) also called as 'Jonschers power law'. From the measure value of imaginary part of dielectric the conductivity data were calculated. Fig. 4 (c) shows the real part of conductivity of our BFN sample at 20, 100, 200 and 300 K temperatures.

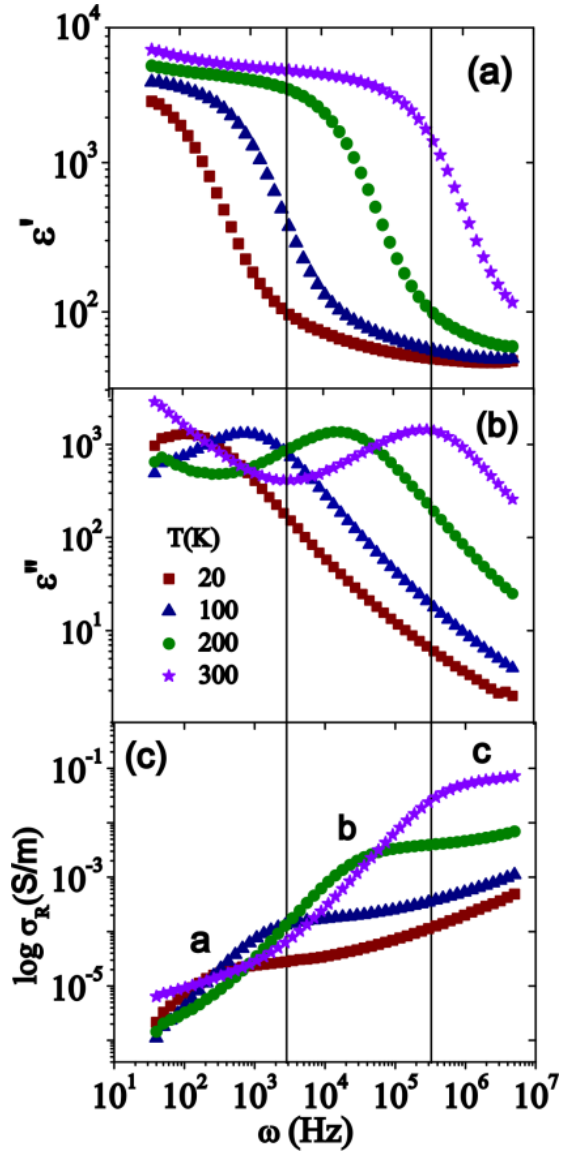

Fig. 4 (a) Real part of dielectric constant, (b) imaginary part of dielectric constant, and (c) real part of conductivity plotted verses frequency of the BFN sample at 20 K, 100 K, 200 K and 300 K temperature; a, b and c shows the different regions of conductivity at 300 K.

From our previous work we had observed that space charge polarization of free and/or bound charge carriers (electrons) and dipolar polarization of bound charge (mainly Fe / Nb ions in O octahedron) are mainly responsible for the high value of dielectric constant observed at low frequency (40Hz)[13]. To understand the contribution of these charges (ions or electrons) to the overall conductivity we had drawn the schematics as shown in Fig. 5. The dash red line mimics the observed overall conductivity of BFN sample, dash dot green line shows the contribution of ions and dash dot dot blue line show the contribution of electrons to the conductivity. .

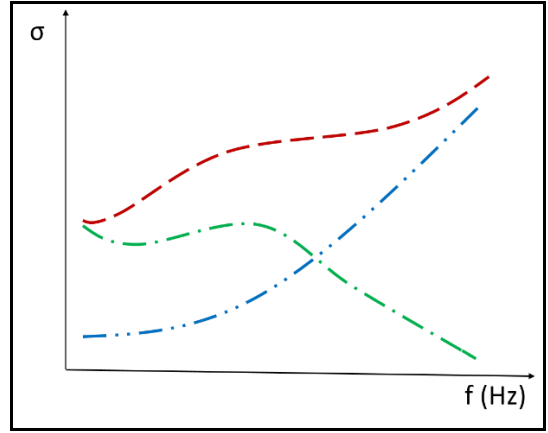

Fig. 5 schematics of overall conductivity of BFN sample with individual contribution of ions and electrons. Red dashed line overall conductivity of BFN, Green dot dash line for ionic conductivity and blue dot dot dash line for electronic conductivity.

There were various reports where observed that, electronic conductivity increases with increased in frequency[29,30,33,34]. This increased in electronic conductivity with increase in infrequency can be understood by considering the random free energy barrier model given by Dyre[34]. According to this model, disorder gives the distribution of energy barriers in the material. At low frequency charge carriers cover a large distances giving very large probability to arrive at high energy barriers, this gives low mobility implies low value of conductivity [$\sigma = q\,n\,\mu$ ($q$: charge, $n$: density of charge carriers, $\mu$: mobility)]. Whereas at high frequency charge carriers able to move for a short distance, gives small probability to arrive at a very large energy barrier giving comparatively large mobility. Therefore high value of conductivity observed at high frequency.

The observed conductivity [Fig. 4 (c)] of our BFN ceramic sample is little complicated having contribution from both ions as wel as electrons. Since, to understand the observed conductivity of our BFN sample we have to consider contribution from the electrons and ions separately. Furthermore, our sample is polycrystalline in nature having grain and grain boundary regions [Fig. 1 (b) and Fig. 2 (e)] contributes differently to the conductivity. Consider conductivity at 300 K which showing three regions marked as **a**, **b** and **c** for frequency regions as, region **a** from 40 - $10^3$ Hz; region **b** between $10^3$ – $6\times10^6$ Hz and region **c** above $10^6$ Hz frequency [Fig. 4 (c)]. The origin of these three regions and there variation with frequency and temperature is discussed as follow.

Considering region **b** where conductivity rises sharply from $3\times10^{-5}$ S/m to $5\times10^{-2}$ S/m between $10^3$ – $6\times10^6$ Hz frequencies range. This observed sharply rise in conductivity is due to the dipolar relaxation[35], relaxation of $Fe^{3+}$ and $Nb^{5+}$ ions with respect to the oxygen octahedra as per our previous work[13]. Since our material is polycrystalline in nature, orientation of dipoles is different for different gains. From the frequency of $10^3$ Hz number of dipoles starts un following the applied electric field (starts relaxing) [Fig. 4(a)]. These relaxing ions with respect to the oxygen octahedra (dipoles) contribute to the conductivity. At a $6\times10^6$ Hz maximum number of dipoles relaxes giving peak maxima in the imaginary part of the dielectric constant [Fig. 4(b)] and corresponding high value of ionic conductivity $5\times10^{-2}$ S/m [Fig. 4(c)]. Above $6\times10^6$ Hz due to decreasing number of dipoles the contribution of ionic conductivity decreases [clearly seen for 200, 100 and 20 K in Fig. 4(b)]. This explains the observed sharply rising conductivity from $3\times10^{-5}$ S/m to $5\times10^{-2}$ S/m observed between $10^3$ – $6\times10^6$ Hz frequencies range. Further as temperature decreases [200, 100 and 20 K in Fig. 4(b)] motion of ions within oxygen octahedron start freezing this leads to the shifting of dipole relaxation and hence ionic conductivity toward lower frequency end as observed in Fig. 4(c) for 200, 100 and 20 K temperatures. The activation energy of this ions were calculated form our previous work and is obtained as 17 meV[13].

The observed conductivity for **a** and **c** regions [Fig. 4(c) for 300 K] can be understand by considering the schematics as shown in Fig. 5 [electronic relaxation]. From our TEM and UV-Vis analysis we had confirmed that there exist the defects at grain and grain boundary region which gives the charge carriers (electrons) for conductivity. Since, the observed low conductivity value ($6\times10^{-6}$ S/m for 300 K) at 40 Hz for region **a**, is due to the electrons at grain boundary region. At low frequency (large period time) these electrons easily conduct through the continues path provided by the grain boundary region, and reached at the other end of the electrode and contribute to the dc conductivity. Meanwhile electrons in the grain regions has the very less (nearly negligible) continues path between electrode available for conduction, therefore there contribution to the conductivity is very less, nearly negligible. As frequency increases the contribution of electrons from grain region progressively start increasing because of the increasing mobility of charge carriers[34]. Meanwhile, the contribution from the grain boundary electrons decreases, almost neglected at high frequency. The increased conductivity of electrons in grain region were not observed [Fig. 4(c) for 300 K] between frequency ranges $10^3 - 6\times10^6$ Hz due to dominated ionic conductivity (as explained already). The region **c** shows nearly constant value of conductivity $5\times10^{-2}$ S/m above $6\times10^6$ Hz frequency [Fig. 4(c) for 300 K]. This is because of the simultaneous contribution of ions and electrons to the conductivity. Above $6\times10^6$ Hz frequency the contribution of ionic conductivity starts decreasing [Fig. 4(b) for 300 K] whereas, the contribution of electronic conductivity increases giving overall constant value of conductivity. The electronic conductivity of grain and grain boundary regions explained in sections 3.5 and 3.4 in detail respectively.

From all these discussion it is clear that in our BFN sample the observed conductivity over measured frequency range is due to the electrons at grain boundary region and ions and electrons of grain region which respond with different frequencies. For 300 K the low value of conductivity observed for the region **a** is due to electrons of grain boundary; region **b** due to ions in the grain region and region **c** due to the combine electrons and ions in the grain region. As temperature decreases from 300 K to 20 K these charge carries start freezing this lead to increase in there relaxation time, resulting in the shifting of overall conductivity toward lower frequency [see Fig. 4(c)].

**3.4 : Electrical conductivity at grain boundary region**

As observed in the HRTEM image of grain boundary region of our BFN sample Fig.2 (f) the ions are randomly arranged. This ions has unsatisfied dangling bond, which has trapped electrons at that lattice site. This trap electrons in the external applied electric field moves from one lattice site to another and produces dipoles (bound charges) or space charge polarization. The motion of these electrons from one lattice site to another gives the conductivity. In presence of external applied electric field some of the electrons can move to the longer lattice site (large path) and some of them can shorter lattice site (short path), its depend on presence of obstacles in their path. At lower frequency (say 40 Hz) very few electrons were able to overcome the obstacle (since by taking larger path) and reached at the other end of the electrode. As frequency increases (say 1000 Hz) number of electrons which having the shorter path increase giving increased conductivity. This explain the increased in conductive of (within defects) grain boundary region with increase in frequency as observed in Fig. 3 (a) for 300 K (region **a**) temperature. Further, from the analysis of SEM [Fig.1 (b)] and TEM [Fig.2 (e)] images the grain boundaries acquired very small region of the pellet as compare to grains. Since conductivity observed at lower frequency (40Hz) is mostly contributed by the grain boundary. As frequency increases contribution from the grain boundary not that much significant as compare to the grain, since neglected at higher frequency. Further as temperature increases (say 400 K) electrons start getting more thermal energy from the lattices (or ions where they were trapped). With increased energy they start responding quickly to electric field, since their response time period to the electric field increased. This result in the shifting of conductivity toward higher frequency [flat region **a** of Fig. 4 (a) for 400 K]. Because of the more numbers of electrons were able to overcome the obstacles occur in their path and gives the increase value of conductivity. This explained the observed increased value $1\times10^{-4}$ S/m (at 40 Hz) for 400 K temperature as compare to $6\times10^{-6}$ S/m for 300 K temperature [Fig. 5 (a)]. To calculated the activation energy of electrons the conductivity data measured between 300 to 400 K temperature ranges [Fig. 6 (a)] were used. The activation energy calculated by using Arrhenius equation given as,

$$\sigma = \sigma_0 \exp\left(-\frac{W}{kT}\right) \quad (6)$$

Where $\sigma_0$ is constant and *W* is the activation energy. Therefor in our BFN material electrons has the activation energy of 317 meV, as shown in Fig.6 (b).

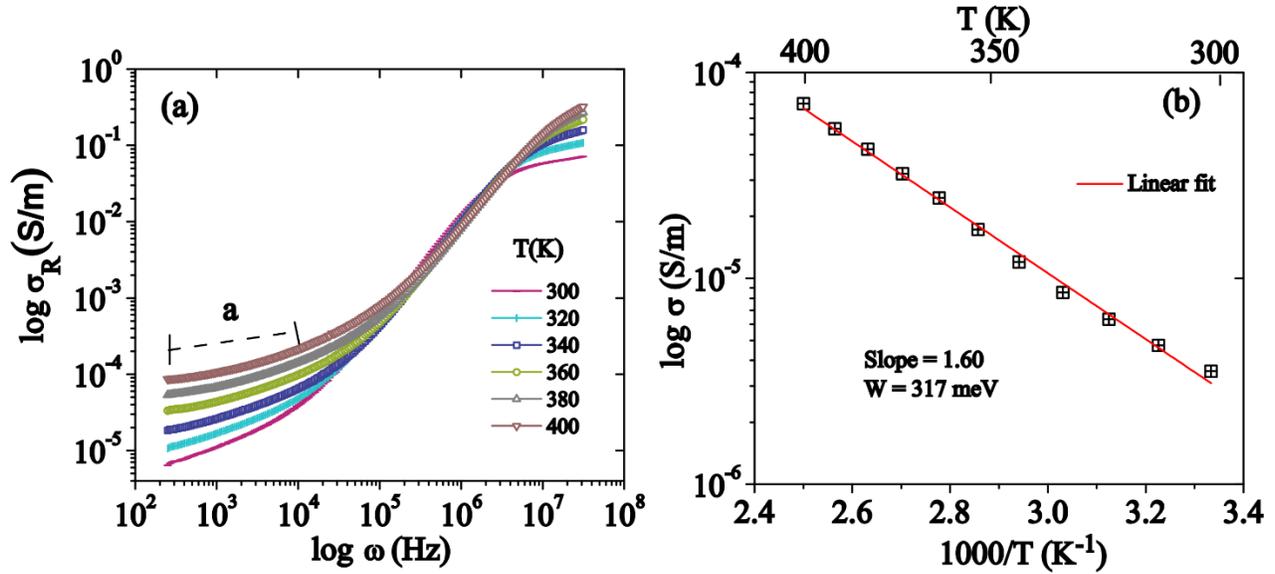

Fig. 6 (a) the conductivity of BFN polycrystalline sample measure between 20 to 400 K. (b) Activation energy calculated between 300 to 400 K temperature range.

### 3.5 : Electrical conductivity in the grain and polarons

In grain regions due to the presence of point defects electrons gets trapped at lattice site. One of the point defect, interstitial were observed in HRTEM image, Fig.2 (g). Also there are reports where presence of oxygen vacancy gives the electrons[20]. All these point defect whether intestinal or vacancies gives the trapped charge carriers at lattice site. Fig. 7 (a) show the electric conductivity of grain region measured between 20 to 300 K temperatures. As per our previous discussion as frequency increases ionic conductivity decrease and electronic conductivity increases simultaneously [regions **c** of Fig. 4(c) for 300 K] giving nearly constant value of conductivity. As frequency increases beyond the $10^7$ Hz electronic part of the grain more dominate, which not able observed due to limited frequency $10^7$ Hz range. But it is clearly observable at 20 K.

Considering conductivity data of 20 K as shown in the Fig. 7 (b), it was observed that for conductivity has value $2\times10^{-5}$ S/m at $10^4$ Hz, further progressively increases with increase in frequency and reach to $4\times10^{-4}$ S/m at $5\times10^7$ Hz frequency. This increased in value of conductivity is due to the increased number of electrons which have shorted path for conduction similarly as explain for grain boundary conduction (section 3.4). Since, as frequency increases mobility of electrons increases giving increased value of conductivity. Also as temperature increases the conductivity value increases and conductivity curve start shifting toward higher frequency [Fig. 7 (b), (c) and (d)]. This is due to the increased energy of electrons with temperature, similarly as explained for grain boundary. As temperature increases electrons starts responding relatively quickly to the applied electric field, shifting conductivity curve toward higher frequency. The increase conductivity value with temperature is due to the increased number bound electrons at grain region.

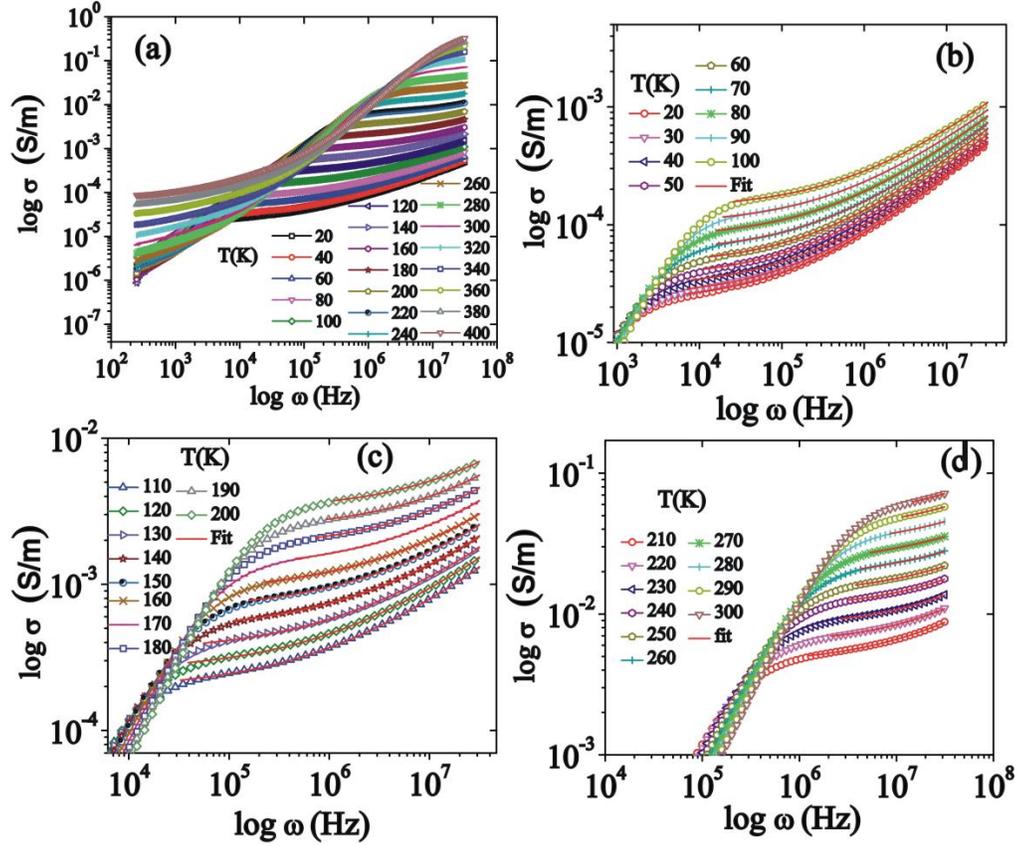

Fig. 7 Conductivity of grain region fitted with Jonscher's power law.

As electron moving within defect (lattice site), due to the coulombic interaction between electrons and surrounding ions they form the quasi particle called as 'polaron'. To study the polaronic conductivity in our BFN sample, conductivity data fitted with Jonscher's power law (Eq.5) and value of conductivity ($\sigma$) and frequency exponent (s) obtained for further analysis[36]. In literature various models are reported to explain the polaronic conduction. Among them barrier hopping (CBH), quantum mechanical tunneling (QMT), small polaron tunneling (SP) and overlap large –polaron tunneling (OLP) models are widely used[29,30]. These models mostly classified on the basis of their behavior of frequency exponent parameter (s) with temperature and frequency. The frequency exponent parameter is obtained by fitting conductivity data by Jonscher's equation (Eq. 5). This parameter gives the rate of change of conductivity with frequency. It is the bridge between observed experimental values to the value calculated by considering theoretical model. Looking at temperature dependent behavior of frequency exponent we can decide which model is preferable to explain the polaronic behavior in the material.

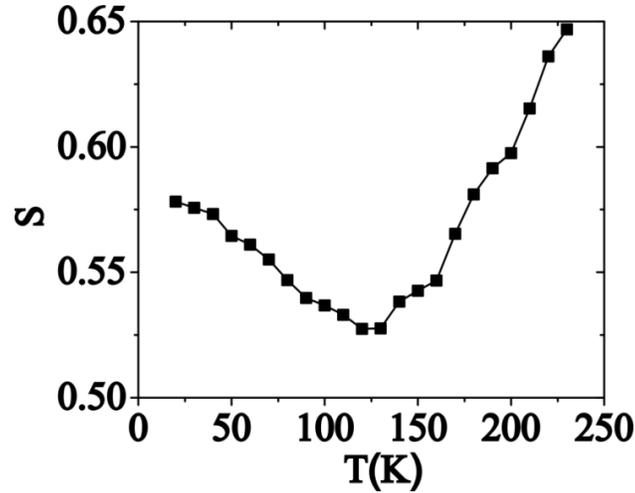

Fig. 8 Variation of frequency exponent (*s*) with temperature

The Fig. 8 shows the temperature dependent frequency exponent (*s*) value of our BFN sample. It is observed that all the values of *s* are less than 1 indication conduction of charge carries occurs between localized sates[32,34]. At 20 K value of *s* is 0.58 which gets reduced till 120 K with 0.54 and further increase with temperature and reached to 0.65 at 240 K. As per temperature depended behaviors of frequency exponent observed in our BFN sample we anticipate that, our material follows overlap large polaron tunneling (OLP) model. According to this model special extend of polaron is large compare to the interatomic spacing and conduction takes place by tunneling. This dielectric or Fröhlich type of conduction are the characteristics of most of the ionic lattices (solid)[30,37]. As observed frequency exponent behavior in our sample, similar type of frequency exponent behavior was also obtained in single crystal $La_2CuO_4$ sample[33]. According to the OLP model the temperature dependence of frequency exponent (*s*) is

given by[29,33]

$$s = 1 - \frac{8\alpha R + 6\beta W(r_0/R)}{[2\alpha R + 6\beta W(r_0/R)]^2} \quad (7)$$

where, $\beta = 1/(k_B T)$, W barrier height for the infinite site separation, $r_0$ polaron radius, $\alpha$ spatial extent of the localized state wave function and $R$ is the hopping distance. Since, temperature dependent frequency exponent given by Eq. 7 has nearly same nature as observed in our sample (Fig. 8), therefore we predict that conductivity of our sample can be explain by the overlap large polaron (OLP) model, which we will validate further.

### 3.6 : Motts variable range hopping (VRH)

According to formulation done in quantum mechanics, for the system, where Fermi energy lies in the range of energies where states are localized (defects states), the hopping conductivity is given as[26],

$$\sigma = 2e^2 R^2\, \upsilon_{ph} N(E_F) exp(-2\alpha R - W/k_B T) \quad (8)$$

Where, $R$ is the electron hopping distance in the direction of the field, $\upsilon_{ph}$ is the phonon frequency, $N(E_F)$ density of defects states within range $k_B T$ of the Fermi energy, $\alpha$ is the rate determining factor at which wave function on a single well falls off with distance (1/ $\alpha$ is the localization length), $W$ difference between two energy levels, and other notations follows their usual meaning. This equation gives relation of the conductivity.

Depending on the average hopping distance $(R_0)$ and decay rate of wavefunction ($\alpha$) with temperature, the electrons in the defects states can have nearest-neighbor hopping (NRH) and Motts variable-range hopping (VRH). Here hopping term used for electrons as for the historic reason, actually electrons are tunneling while ions are hopping between defect states[34]. The NRH expected highly localized defect states (Anderson localized) where $\alpha R_0 \gg 1$, and conductivity follows $exp(-W/kT)$ type Arrhenius behavior (Eq. 6). Furthermore, in VRH, there is enough spread of Anderson localized wavefunction compare to the average hopping distance, where $\alpha R_0$ comparable or $< 1$. Due to overlap of the wave function hopping distance increases with decreasing temperature (hopping range varies with temperature). The corresponding conductivity follows $exp(-B/T^{1/4})$ behavior[38,39]. Conversely by observing behavior of conductivity with respect to the exponential power dependent of temperature (either T or $T^{1/4}$), one can identify NRH and VRH conductivity mechanism in defect states of the material.

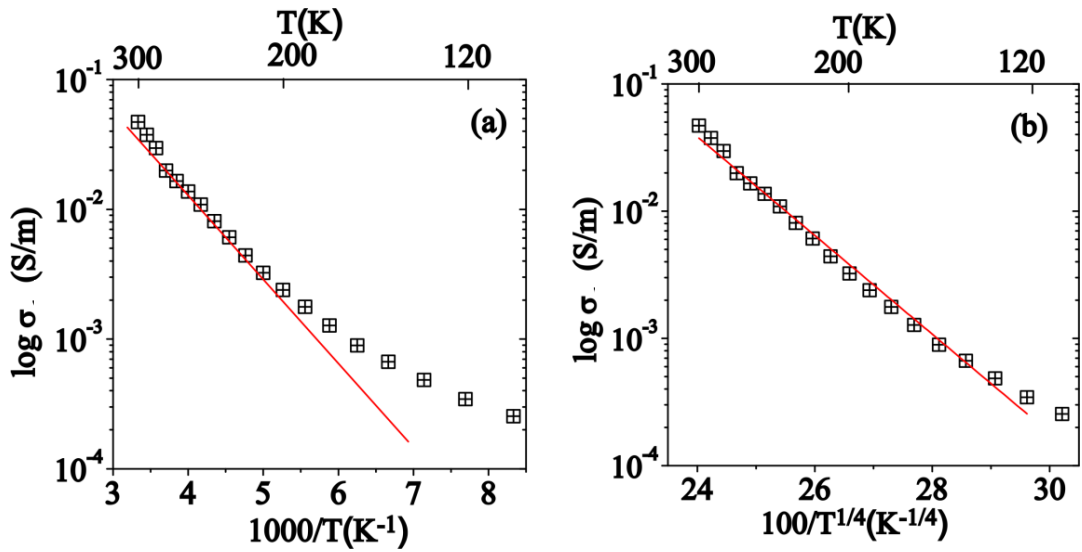

Fig. 9 (a) Arrhenius plot (1000/T) showing deviation from NRH mechanism (b) VRH plot $(100/T)^{1/4}$ showing the linear behavior.

The Fig.9 (a) and (b) shows the conductivity plot for 1/T and $1/T^{1/4}$ respectively. From the plots it is observed that in 1/T (Fig.9 (a)) conductivity for measured temperature range deviate from the linearity, whereas in $1/T^{1/4}$ it almost linear (Fig.9 (a)). This shows that conductivity of our material follows VRH mechanism. Further to calculate activation energy and hopping distance of polaron the formulation given by the VRH model used. According to the VRH model the temperature dependent conductivity given as,[26,36,38]

$$\sigma = \sigma_0 exp\,[-(T_0/T)^{1/4}] \quad (9)$$

$$\text{Where, } T_0 = 24/[\pi k_B N(E_F)\xi^3] \quad (10)$$

$$\text{Hopping distance } R = \xi^{1/4}/\{8\pi N(E_F)k_B T\}^{1/4} \quad (11)$$

$$\text{\& Activation energy } W = 3/4\pi R^3 N(E_F) \quad (12)$$

All the parameters has usual meaning as explain previously. The conductivity data (Fig.9 (b)) fitted with Eq. 9 and corresponding parameters were obtained. The obtained value for parameters $\sigma_0$ and $T_0$ given as $7.81 \times 10^7$ S/cm and $2.27 \times 10^6$ K respectively. By using Eq. 10 calculated value of density of defect states $N(E_F)$ comes to be $3.17 \times 10^{17}$ eV$^{-1}$cm$^{-3}$. Here for the calculation of $N(E_F)$ we had choose localized length $\xi = 1/a = 5$ nm ($a$ is the lattice constant). From our previous work obtained value of lattice constant was 0.4 nm[13]. There are various reports on value of the localized length, in range of 5-6 nm for large polaron by DFT calculation[40,41], which further validate our choice for localized length. By knowing the value of $N(E_F)$ for particular temperature average hopping distance $R$ and activation energy $W$ is calculated by using Eq. 11 and Eq. 12 respectively. We calculated the average hopping distance and activation energy for various temperatures. The calculated value $R$ and $W$ for various temperatures plotted in Fig. 10 (a) and (b) respectively.

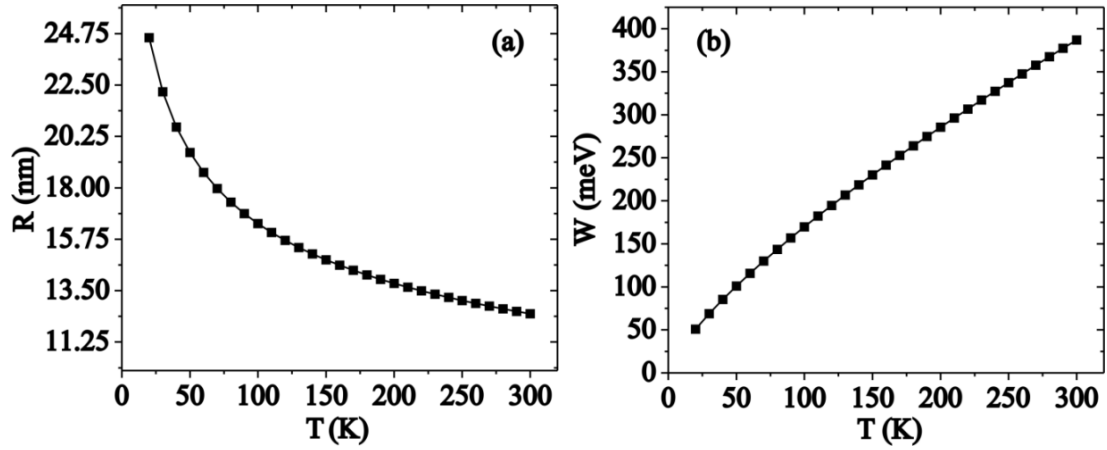

Fig. 7 (a) shows the values of hopping range R and (b) activation energy W with temperature.

From Fig.10 (a) it is observed that average hopping distance at 300 K is 12.48 nm with decreasing temperature it progressively increases and reached 24.57 nm for 20 K temperature, shows that with decreasing temperature hopping range increases. Further from Fig.10 (b) it is observed that activation energy at 300 K is 387 meV, which decreases with decreasing temperature and reached 50 meV at 20 K temperature. Which is expected as temperature decreases energy required for carriers to overcome the potential barrier is decreases hence they can move to the longer distance at lower temperature. Which is the characteristics of Mott VRH model[26,38,39]. If we compare the energy of electrons calculated by purely due to temperature i.e. $k_BT$ for 20 and 300 K it is obtained to be 1.72 meV and 25.83 meV respectively with the energy acquired by the electrons obtained by above calculation 50 meV (for 20 K) and 387 meV (for 300 K), it is observed that energy of electrons is high. With this high energy electron can easily tunnel through the potential barrier and reached at larger distances (Fig.10 (a)). This further confirmed that observed conductivity in our BFN sample is occurred due to the electrons (polaron) tunneling between the defect states.

## 3.7 : Phonon contribution to the conductivity

As given in the Eq. 7, there is contribution of phonon to the conductivity. To understand contribution of phonons to the conductivity of our BFN sample, the conductivity data from 20 K to 300 K temperature range is plotted, as shown in Fig.11. There observed to be two different linear regions first is between 300 K to 120 K and second is between 120 K to 20 K temperature range. The corresponding activation energy of electrons were calculated by using Arrhenius equation as $110 \pm 2$ meV and $1.6 \pm 0.3$ meV for first and second region respectively. These observed two different activation energy of electron with different region can be understand by Schnakenberg model of conductivity of polaronic impurity hopping[42,43]. According to this model, as temperature decreases contribution of various phonons to the conductivity start decreasing progressively. For a temperature above $\Theta_D/2$ polaronic transport is assisted by the multiphonons, acoustic as well as optical phonons. Whereas below $\Theta_D/2$ polaronic transport assisted by the acoustic phonons only. Here $\Theta_D$ is the Debye temperature.

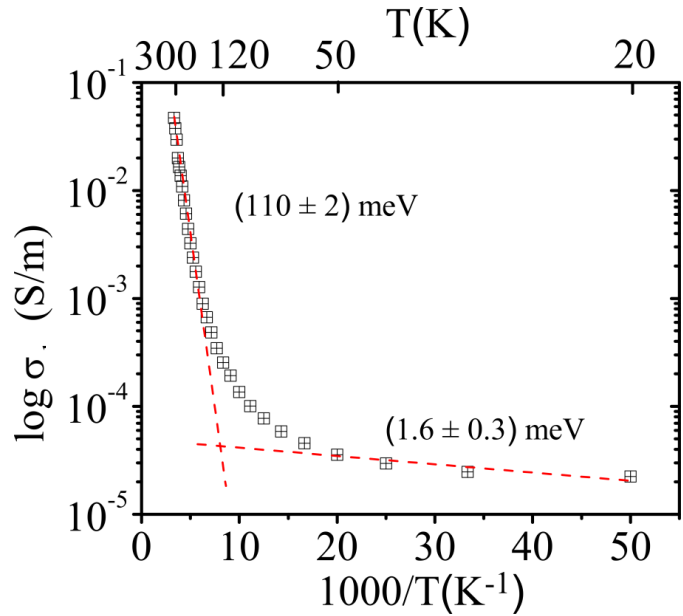

Fig. 8 Contribution of phonon to the conductivity with temperature.

Therefore, in our BFN sample observed high activation energy 110 ± 2 meV for first region is due to the contribution of acoustic and optical phonons (multiphonons). Whereas observed energy value 1.6 ± 0.3 meV for second region is due to contribution of acoustic phonons only. As temperature decreases contribution from the different phonons to the conductivity starts decreasing.

From the above discussions it is confirmed that, conductivity of grain region of our BFN sample is due to the polarons. According to Mott variable hopping model (VRH) observed in our sample, electrons tunnel between the defect states. And also the behavior of frequency exponent with temperature we confirmed that our material follows overlap large polaron (OLP) tunneling model of conductivity.

**3.8: Origin of charge carrier due to intrinsic defects in BFN ceramics**

Since our BFN ceramic sample synthesized by solid state route at an 1200ºC, formed powder is in polycrystalline nature. Further synthesized powder were pressed into the pallet and sintered at high temperature in air. Due to all these processes there are various kinds of defects present in our sample[20,44]. This point defects can be vacancies, interstitial (observed in HRTEM) etc. These defect gives the charge carriers for conduction. In our BFN sample few possible defects represented in kröger-Vink notation as,

Oxygen vacancy at site

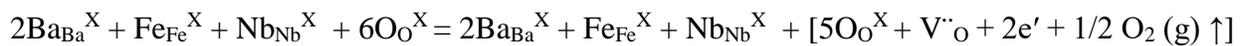

$2Ba_{Ba}^X + Fe_{Fe}^X + Nb_{Nb}^X + 6O_O^X = 2Ba_{Ba}^X + Fe_{Fe}^X + Nb_{Nb}^X + [5O_O^X + V_O^{\cdot\cdot} + 2e' + 1/2\ O_2\ (g) \uparrow]$

Oxygen interstitial at site

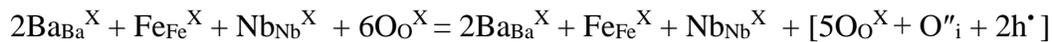

$2Ba_{Ba}^X + Fe_{Fe}^X + Nb_{Nb}^X + 6O_O^X = 2Ba_{Ba}^X + Fe_{Fe}^X + Nb_{Nb}^X + [5O_O^X + O_i'' + 2h^{\cdot}]$

Metal ion vacancy and interstitial / Frenkel defect (only shown for Fe)

$$2Ba_{Ba}^X + Fe_{Fe}^X + Nb_{Nb}^X + 6O_O^X = 2Ba_{Ba}^X + [\,V'''_{Fe} + 3h^\bullet + Fe_i^{\bullet\bullet\bullet} + 3e'\,] + Nb_{Nb}^X + 6O_O^X$$

All these created intrinsic defects gives the charge carrier for the conduction in our BFN sample. For example, electrons occur due to the oxygen vacancy get absorbed by the $Fe^{3+}$ ion and get converted into the $Fe^{2+}$, this electron move between Fe lattice site and contribute to the conductivity. $Fe^{3+} + e' = Fe^{2+}$

## 4. Conclusion:

Low temperature ac conductivity measurement done on the dense polycrystalline BFN pellet between temperature ranges of 20 to 300K. The pXRD and SEM images shows the phase pure polycrystalline sample having grain and grain boundary region. Phase purity of sample further confirmed by the HRTEM and SAED pattern. Further in HRTEM image there observed to be intrinsic point defect at grain and surface defect at grain boundary region. Existence of various defects present in our sample further confirmed by calculating Urbach energy (534 meV) associated with the defects level (band gap energy 1.88 eV). In presence of external applied electric field electrons trapped at the defects contribute to electronic conductivity at grain and grain boundary region along with the ionic conductivity of the grain region. At lower frequency electronic conductivity of grain boundary; at intermediate frequency ionic conductivity of grain and at higher frequency electronic conductivity of grain region dominated (for 300 K). As temperature decrease (below 300 K) due to freezing motion of this charge carries, conductivity start shifting toward lower frequency. The activation of electrons at grain boundary region and ions at grain region calculated to be 317 meV and 17 meV respectively.

The electronic conductivity of grain obeys the jonscher's power low. The temperature dependent frequency exponent shows that overlap large polaron model (OLP) is suitable to explain the observed conduction of grain. Temperature dependent conductivity follows the Mott variable range hopping model (VRH) indicating that defect are randomly distribute, with defect density of $3.17 \times 10^{17}$ eV$^{-1}$cm$^{-3}$. The phonons contribution to the electronic conductivity explained by considering Schnakenberg model, where observed activation energy of 110 meV corresponds both optical and acoustic phonons, and 1.6 meV corresponds to the acoustic phono only. All these study confirmed that the charge carrier at grain regions are the polaron and they tunnel between the defects states. Thus our work provide the systematic study of the low temperature conductivity mechanism of polycrystalline BFN ceramics.

**Conflicts of Interest**

The authors declare no conflicts of interest.

**Reference:**

[1] Sung-yoon chung, I.-D. Kim, Suk-Joong, and L. Kang, Nat. Mater. **3**, 774 (2004).

[2] X. Hao, J. Adv. Dielectr. **3**, 1330001 (2013).

[3] S. Matteppanavar, S. Rayaprol, B. Angadi, and B. Sahoo, J. Alloys Compd. **677**, 27 (2016).

[4] S. Matteppanavar, S. Rayaprol, A. V. Anupama, B. Angadi, and B. Sahoo, Ceram. Int. **41**, 11680 (2015).

[5] S.T. Dadami, S. Matteppanavar, I. Shivaraja, S. Rayaprol, B. Angadi, and B. Sahoo, J. Magn. Magn. Mater. **418**, 122 (2016).

[6] W. Jiagang, X. Dingquan, and J. Zhu, Chem. Rev. **115**, 2559 (2015).

[7] H.S. Mohanty, T. Dam, H. Borkar, D.K. Pradhan, K.K. Mishra, A. Kumar, B. Sahoo, P.K.

Kulriya, C. Cazorla, J.F. Scott, and D.K. Pradhan, Jounral Phys. Condens. Matter **31**, 1 (2019).

[8] S. Madolappa, A. V. Anupama, P.W. Jaschin, K.B.R. Varma, and B. Sahoo, Bull. Mater. Sci. **39**, 593 (2016).

[9] S. Vasala and M. Karppinen, Prog. Solid State Chem. **43**, 1 (2015).

[10] N. Setter and L.E. Cross, J. Appl. Phys. **51**, 4356 (1980).

[11] D. Viehland, S.J. Jang, L.E. Cross, and M. Wuttig, J. Appl. Phys. **68**, 2916 (1990).

[12] V. V. Shvartsman and D.C. Lupascu, J. Am. Ceram. Soc. **95**, 1 (2012).

[13] V. Khopkar and B. Sahoo, Phys. Chem. Chem. Phys. **22**, 2986 (2020).

[14] Saha and T.P. Sinha, Phys. Rev. B **65**, 1 (2002).

[15] S. Saha and T.P. Sinha, J. Phys. Condens. Matter **14**, 249 (2002).

[16] I. P. Raevski, S.A. Prosandeev, A.S. Bogatin, M.A. Malitskaya, and L. Jastrabik, J. Appl. Phys. **93**, 4130 (2003).

[17] Uraiwan Intatha, S. Eitssayeam, J. Wang, and T. Tunkasiri, Curr. Appl. Phys. **10**, 21 (2010).

[18] M. Ganguly, S. Parida, E. Sinha, S.K. Rout, A.K. Simanshu, A. Hussain, and I.W. Kim, Mater. Chem. Phys. **131**, 535 (2011).

[19] S. Bhagat and K. Prasad, Phys.Status Solidi A **207**, 1232 (2010).

[20] Z. Wang, X.M. Chen, L. Ni, and X.Q. Liu, Appl. Phys. Lett. **90**, 1 (2007).

[21] P. Kubelka, J. Phys. Condens Matter **31**, 1 (1931).

[22] V. Džimbeg-Malčić, Ž. Barbarić-Mikočević, and K. Itrić, Teh. Vjesn. **19**, 191 (2012).

[23] J. Tauc, Mater. Res. Bull. **5**, 721 (1970).

[24] J. Tauc, Mat. Res. Bull. Vol. **3**, 37 (1968).

[25] O. Stenzel, *The Physics of Thin Film Optical Spectra*, Second (Springer Berlin Heidelberg, 2005).

[26] N.F. Mott and E.A. Davis, *Electronic Processes in Noncrystalline Materials*, Second (Oxford University press Inc, New York, 1979).

[27] G.E. Pike, Phys. Rev. B **6**, 1572 (1972).

[28] M. Pollak and T.H. Geballe, Phys. Rev. **122**, 1742 (1961).

[29] S.R.Elliott, Advanes Phys. **36**, 135 (1987).

[30] A.R. Long, Advanes Phys. **31**, 553 (1982).

[31] A.K. Jonscher, *Dielectric Relaxation in Solids* (Chelsea Dielectric Press, London, 1983).

[32] G.G. Raju, *Dielectrics in Electric Field* (Marcel Dekker, Inc, New York, 2003).

[33] P. Lunkenheimer, G. Knebel, A. Pimenov, G.A. Emel'chenko, and A. Loidl, Zeitschrift Für Phys. B Condens. Matter **99**, 507 (1996).

[34] J.C. Dyre, J. Appl. Phys. **64**, 2456 (1988).

[35] J. van Heumen, W. Wieczorek, M. Siekierski, and J.R. Stevens, J. Phys. Chem. **99**, 15142 (1995).

[36] L. Zhang and Z. Tang, Phys. Rev. B **70**, 174306:1 (2004).

[37] H. Fröhlich, Adv. Phys. **3**, 325 (1954).

[38] N. F. Mott, Philos. Mag. **17:150**, 1259 (1968).

[39] N.F. Mott, J. Non- Crstalline Solids **8–10**, 1 (1972).

[40] K. Miyata, D. Meggiolaro, M.T. Trinh, P.P. Joshi, E. Mosconi, S.C. Jones, F. De Angelis, and X.-Y. Zhu, Sci. Adv. **3**, 1701217 (2017).

[41] F. Zheng and L.-W. Wang, Energy Environ. Sci. (2019).

[42] J. Schnakenberg, Phys. Status Solidi B **28**, 623 (1968).

[43] R. Karsthof, M. Grundmann, A.M. Anton, and F. Kremer, Phys. Rev. B **99**, 1 (2019).

[44] C. Ran, J. Xu, W. Gao, C. Huang, and S. Dou, Chem. Soc. Rev. **47**, 4581 (2018).

[45] R.P. Gupta and S.K. Sen, Phys. Rev. B **10**, 71 (1974).

[46] R.P. Gupta and S.K. Sen, Phys. Rev. B **12**, 15 (1975).

[47] S.P. Kowalczyk, L. Ley, F.R. McFeely, and D.A. Shirley, Phys. Rev. B **11**, 1721 (1975).

[48] L. Yin, I. Adler, T. Tsang, L.J. Matienzo, and S.O. Grim, Chem. Phys. Lett. **24**, 81 (1974).

[49] A.P. Grosvenor, B.A. Kobe, M.C. Biesinger, and N.S. McIntyre, Surf. Interface Anal. **36**, 1564 (2004).

# ESI

Fig.1(ESI) shows the wide (survey) spectra of BFN sample with analyzed with Casa XPS software. Table 1(ESI) gives the peak position, area under the curve and % atomic concentration obtained on the surface of the BFN pellet sample. Chemical formula calculated from the % atomic concentration is $BaFe_{0.355}Nb_{0.462}O_{3.3}$ is nearly equivalent to our required $BaFe_{0.5}Nb_{0.5}O_3$. Deviation from the stoichiometric is can be due to the presence of less amount of Fe and Nb elements at surface. Presence of excess amount of O and C is the contamination arises may be from binder (PVA) used in the sample preparation and adventitious carbon. High resolution XPS spectra of Fe cation investigated further to study the presence of defects (see ESI for high resolution XPS data of Ba, Nb, O and C components.

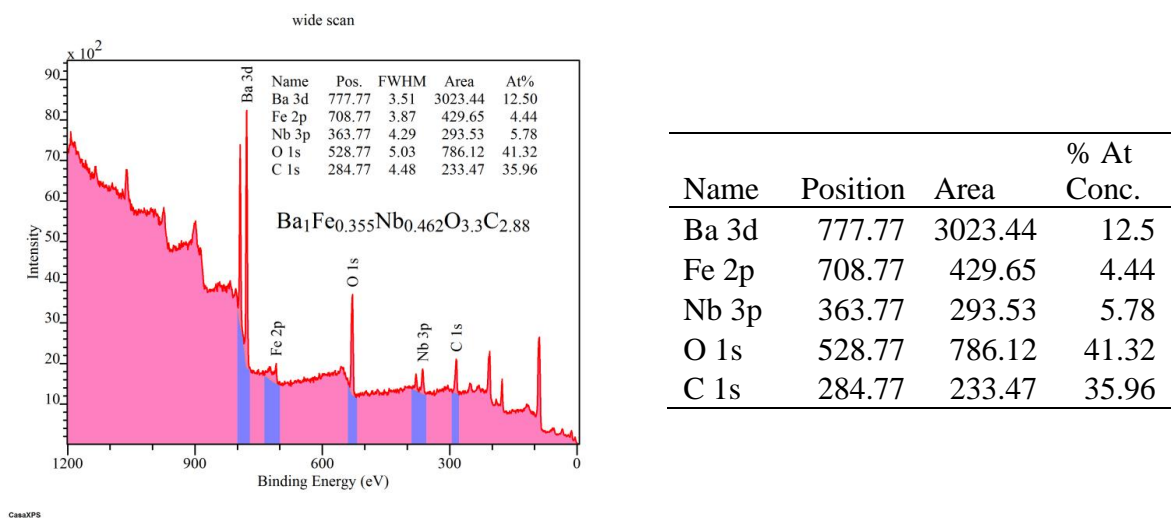

| Name | Position | Area | % At Conc. |
|---|---|---|---|
| Ba 3d | 777.77 | 3023.44 | 12.5 |
| Fe 2p | 708.77 | 429.65 | 4.44 |
| Nb 3p | 363.77 | 293.53 | 5.78 |
| O 1s | 528.77 | 786.12 | 41.32 |
| C 1s | 284.77 | 233.47 | 35.96 |

Fig. 1(ESI): Wide scan spectra of BFN sample, showing presence of % atomic concentration at surface.
Table 1: Obtained peak position, area under the curve and % atomic concentration of composition.

Fig. 2(ESI) shows plot of the Fe $2p_{3/2}$ fitted with the Gupta and Sen (GS) multiplets [45–47] (P1- P4), low BE pre-peak (P5) and satellite peak[48] (satellite). The $2p_{3/2}$ envelopes for BFN compound fit well with the GS multiplet, the corresponding value peak position and % area of four GS peaks along with pre-peak and satellite peak were tabulated (Table 2). The energy difference between peak having value nearly equal as reported in the literature[49]. The multiplets splitting of the Fe $2p_{3/2}$ shows that $Fe^{3+}$ is in the high spin state. Also presence of broad peak without any shoulder shows that in BFN Fe is in the $Fe^{3+}$ state, there is no contribution from the $Fe^{2+}$ at surface. There observed to be low BE peak at 707.52 eV corresponding to the defect in the neighboring site.

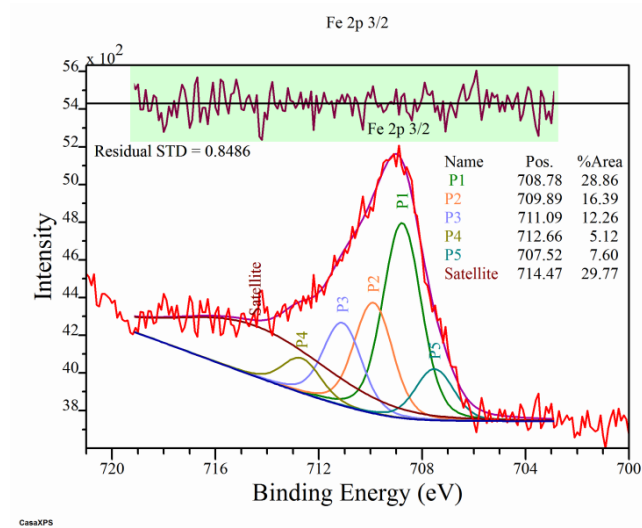

| Name | Position (eV) | % Area |
|---|---|---|
| P1 | 708.78 | 28.86 |
| P2 | 709.89 | 16.39 |
| P3 | 711.09 | 12.26 |
| P4 | 712.66 | 5.12 |
| P5 | 707.52 | 7.6 |
| Satellite | 714.47 | 29.77 |
| P2 - P1 | 1.11 | |
| P3 - P2 | 1.20 | |
| P4 - P3 | 1.57 | |

Fig. 2(ESI): Wide scan spectra of BFN sample, showing presence of % atomic concentration at surface.
Table 2: Obtained peak position, area under the curve and % atomic concentration of composition.

From the analysis of wide scan and high resolution of Fe XPS data it observed that at surface of our BFN sample there is presence of defect.

Fig. 3(ESI) shows the high resolution XPS data of Ba 3d. There observed two peaks with peak position at 778.38 eV and 793.72 eV correspond to the spin orbit coupling (L-S) of Ba 3d 5/2 and 3d 3/2 respectively. The energy difference between these two peaks is the 15.34 eV. Two components were used to fit each peak as shown in the Fig. 1(ESI). This multiplets splitting of Ba 3d 5/2 and 3d 3/2 is due to the presence of different environments (crystal field) for Ba atom.

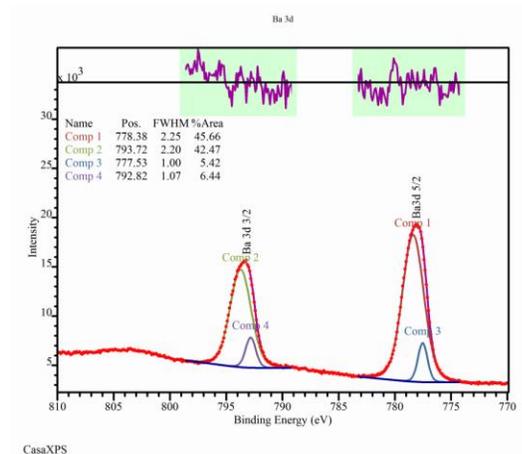

Fig 3 (ESI): High resolution XPS data of Ba 3d with peak model.

Fig 4 (ESI) shows the high resolution XPS data of C 1s. Three components were used to fit the data, these are correspond to the C-C (284.56 eV) C-O (288.02 eV) and C= O (285.60 eV). Presence of adventitious C is can be due to binder (3 wt% Polyvinyl Alcohol) used for sample preparation. This explains the C-C and C-O peak. While at high temperature sintering there is the formation of carbonate this explain the C=O present in the peak model.

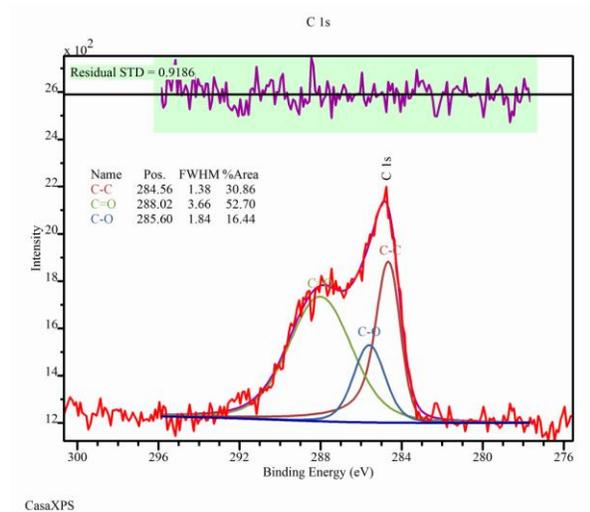

Fig 4 (ESI): High resolution XPS data of C 1s with peak model.

Fig 5 (ESI) shows the high resolution XPS data of O 1s. Four components were used to fit the data, these are correspond to the metal oxide, M-O1 and M-O2 with peak position 529.02 eV and 530.13 eV respectively. These peaks are corresponds to Ba /Fe / Nb bond with O. Also, C-O (531.51 eV) and C= O (532.63 eV) observed these are due to the used of PVA binder and carbonate.

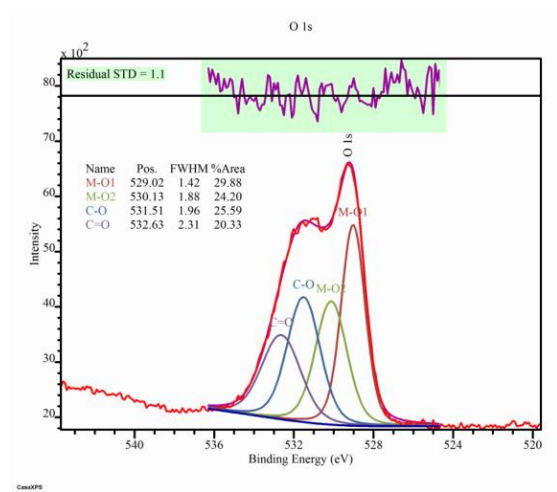

Fig 5 (ESI): High resolution XPS data of O 1s with peak model.

Fig 6(ESI) shows the high resolution XPS data of Nb 3d. There observed two peaks with peak position at 205.76 eV and 207.98 eV correspond to the spin orbit coupling (L-S) of Nb 3d 5/2 and 3d 3/2 respectively. The energy difference between these two levels is the 2.22 eV. Two components were used to fit each peak as shown in the Fig. 4(ESI). This multiplets splitting of Nb 3d 5/2 and 3d 3/2 is due to the presence of different environments (crystal field) for Nb atom. There observed to be two satellites peaks SL-1 and Sl-2 with BE of 213.03 and 209.85 eV respectively.

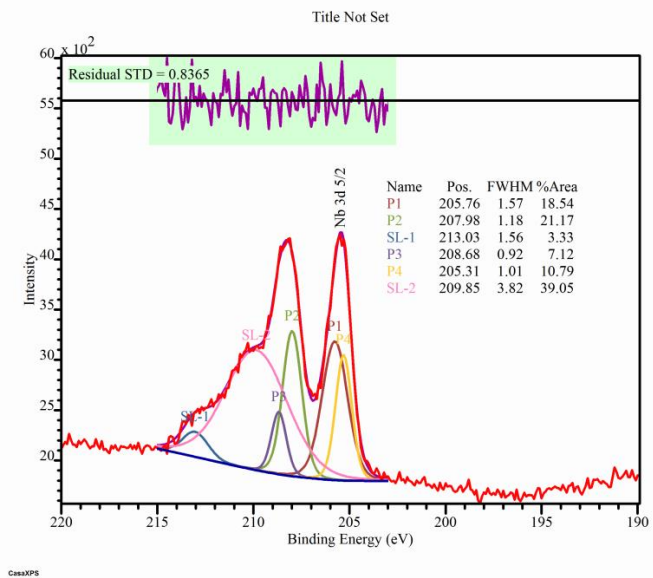

Fig 6 (ESI) :High resolution XPS data of Nb 3d with peak model.

Fig 7 (ESI)shows the high resolution XPS data of Fe 2p. There observed two peaks with peak position at 708.62 eV and 722.59 eV correspond to the spin orbit coupling (L-S) of Fe 2p 3/2 and 2p 1/2 respectively. The energy difference between these two levels is the 13.97eV. Four components were used to Fe 2p 3/2 peak as shown in the Fig. 7(ESI). This multiplets splitting of Fe 2p 3/2 is due to the crystal field interaction between unpaired electrons in 3d energy level with the electron in the 2p 3/2. Due to crystal field interaction degeneracy of 2p 3/2 gets lifted giving four energy levels. Energy of these four levels (P1-P4) is tabulated. The similar explanation is also valid for Fe 2p 1/2 state. There observed two satellite peaks 3/2 SL and 1/2 SL with BE 714.9 and 731.35 eV respectively.

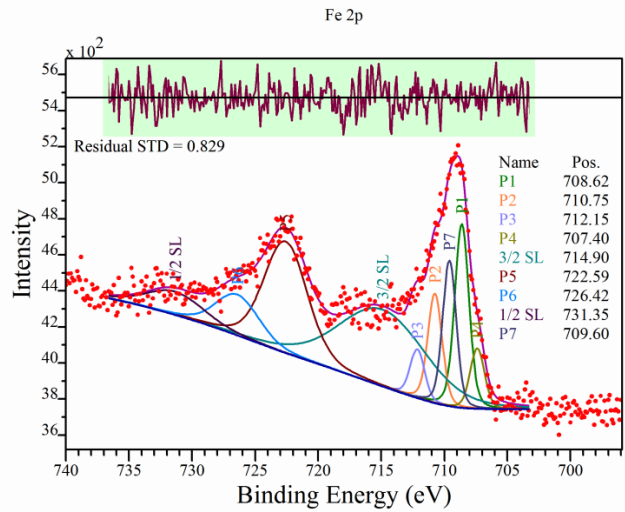

| Name  | Position (eV) | % Area |
|-------|---------------|--------|
| P1    | 708.62        | 8.88   |
| P7    | 709.6         | 6.63   |
| P2    | 710.75        | 5.29   |
| P3    | 712.15        | 2.36   |
| P4    | 707.4         | 2.92   |
| 3/2 SL| 714.9         | 22.53  |
| P5    | 722.59        | 31.34  |
| P6    | 726.42        | 11.46  |
| 1/2 SL| 731.35        | 8.59   |

Fig7 (ESI) :High resolution XPS data of Fe 2p with peak model.

Table 3: Obtained peak position, area under the curve and % atomic concentration of composition.